\documentclass[twocolumn,english,conference]{IEEEtran}
\usepackage[T1]{fontenc}
\usepackage{babel}
\usepackage{float}
\usepackage{amsmath}
\usepackage{amssymb}
\usepackage{graphicx}
\usepackage[unicode=true,
 bookmarks=true,bookmarksnumbered=true,bookmarksopen=true,bookmarksopenlevel=1,
 breaklinks=false,pdfborder={0 0 0},backref=false,colorlinks=false]
 {hyperref}
\hypersetup{pdftitle={Your Title},
 pdfauthor={Your Name},
 pdfpagelayout=OneColumn, pdfnewwindow=true, pdfstartview=XYZ, plainpages=false}
\usepackage{breakurl}

\makeatletter

\floatstyle{ruled}
\newfloat{algorithm}{tbp}{loa}
\providecommand{\algorithmname}{Algorithm}
\floatname{algorithm}{\protect\algorithmname}

\usepackage[caption=false,font=footnotesize]{subfig}

\makeatother

\begin{document}

\title{Signal and Interference Leakage Minimization in MIMO Uplink-Downlink
Cellular Networks}

\author{\IEEEauthorblockN{\IEEEauthorrefmark{1}Tariq~Elkourdi, \IEEEauthorrefmark{1}Osvaldo~Simeone,
\IEEEauthorrefmark{2}Onur~Sahin and \IEEEauthorrefmark{3}Shlomo~Shamai
(Shitz)}\IEEEauthorblockA{\IEEEauthorrefmark{1}CWCSPR, New Jersey Institute of Technology,
Newark, NJ, 07102, USA }\IEEEauthorblockA{\IEEEauthorrefmark{2}InterDigital Inc., Melville, New York, 11747,
USA}\IEEEauthorblockA{\IEEEauthorrefmark{3}Department of Electrical Engineering, Technion,
Haifa, 32000, Israel\\
Email: \{tariq.elkourdi,osvaldo.simeone\}@njit.edu, onur.sahin@interdigital.com,
sshlomo@ee.technion.ac.il}}

\IEEEaftertitletext{after title text like dedication}
\maketitle
\begin{abstract}
Linear processing in the spatial domain at the base stations (BSs)
and at the users of MIMO cellular systems enables the control of both
inter-cell and intra-cell interference. A number of iterative algorithms
have been proposed that allow the BSs and the users to calculate the
transmit-side and the receive-side linear processors in a distributed
manner via message exchange based only on local channel state information.
In this paper, a novel such strategy is proposed that requires the
exchange of unitary matrices between BSs and users. Specifically,
focusing on a general both uplink- and downlink-operated cells, the
design of the linear processors is obtained as the alternating optimization
solution of the problem of minimizing the weighted sum of the downlink
and uplink inter-cell interference powers and of the signal power
leaked in the space orthogonal to the receive subspaces. Intra-cell
interference is handled via minimum mean square error (MMSE) or the
zero-forcing (ZF) precoding for downlink-operated cells and via joint
decoding for the uplink-operated cells. Numerical results validate
the advantages of the proposed technique with respect to existing
similar techniques that account only for the interference power in
the optimization.\end{abstract}

\section{Introduction}

Linear processing in the spatial domain at the base stations (BSs)
and at the users of a Multi-Input Multi-Output (MIMO) cellular system
is a well studied technique that enables the control of both inter-cell
and intra-cell interference (see, e.g., \cite{Shi}). A number of
iterative algorithms have been proposed in the past few years for
the design of the linear processors that are either centralized, see,
e.g., \cite{Kaviani} and references therein, or can be instead implemented
in a decentralized way \cite{Shi}\cite{Jafar}-\cite{Schreck}. In
the latter case, the BSs and the users calculate the transmit-side
and the receive-side linear processors in a distributed manner via
message exchange based only on local channel state information. 

The distributed techniques in \cite{Shi}\cite{Jafar}-\cite{Schreck}
differ in the information that is exchanged between the BSs and users
and in the processing that is carried out at the two sides. Another
key classification of these techniques can be done with respect to
methods that apply to MIMO interference channels, i.e., cellular systems
with a single user per cell, and techniques are suitable for to more
general cellular systems with multiple users per cell. The interference
leakage minimization (ILM) techniques of \cite{Jafar}\cite{Peters}
require the exchange of unitary matrices between the two sides%
\footnote{A different implementation based on pilot symbols and estimation is
also possible, see \cite{Jose}.%
} and was proposed for a MIMO interference channel. References \cite{Zhuang}\cite{Schreck}
generalize the ILM technique to a cellular system with an arbitrary
number of users per cell, where the cells operate in either uplink
or downlink. In contrast, the technique proposed in \cite{Shi} requires
the exchange of additional information beside unitary matrices and
applies to the downlink of a general MIMO cellular system. In this
regard, we observe that the transmission of unitary matrices is facilitated
by the advances in the quantization over the Grassmann manifold (see,
e.g., \cite{Grassman}) and is hence desirable, making the ILM scheme
of \cite{Jafar}\cite{Peters}\cite{Zhuang}\cite{Schreck} potentially
more viable for practical implementation. The signal plus interference
leakage minimization technique (SILM) of \cite{Kumar} modifies the
ILM strategy by including in the cost function, not only the interference
power, but also the power of the signal that is wasted in the space
orthogonal to the receive subspaces. This scheme also requires the
exchange of unitary matrices and was studied in \cite{Kumar} for
MIMO interference channels.

\begin{figure}[t]
\centering\includegraphics[clip,scale=0.5]{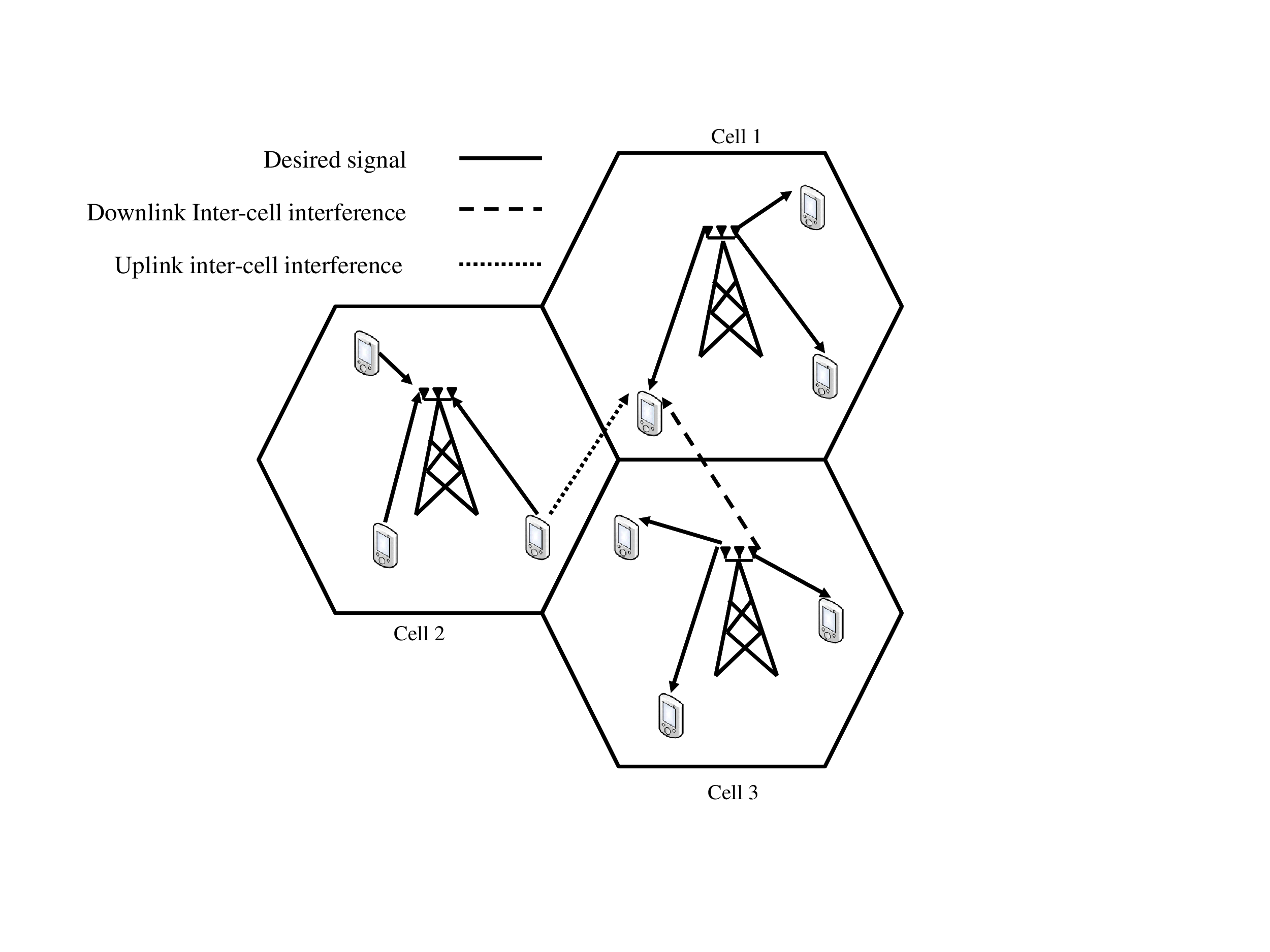}\protect\caption{\label{fig:sys_model} Multi-cell uplink-downlink MIMO system. Downlink
and uplink inter-cell interference signal paths are shown for a user
in cell 1 as an example.}
\end{figure}

In this paper, a novel iterative strategy is proposed that generalizes
SILM \cite{Kumar} to a MIMO cellular system with an arbitrary number
of users per cell and in which each cell may be operated either in
the uplink or in the downlink. Note that this mixed uplink-downlink
configuration is known to be potentially advantageous, even in terms
of degrees of freedom \cite{Jeon}. Specifically, following \cite{Zhuang}\cite{Schreck}\cite{Tse},
the precoding matrix at the downlink-operated BSs is factorized into
a unitary pre-processing matrix that handles inter-cell interference
and a post-processing matrix that deals with intra-cell interference.
The design of precoding and receiver-side matrices is obtained as
the alternating optimization solution of the problem of minimizing
the weighted sum of the inter-cell interference powers and of the
signal power leaked in the space orthogonal to the receive subspaces.
The post-processing precoding matrices at the downlink BSs are then
calculated using the minimum mean square error (MMSE) or the zero-forcing
(ZF) criteria, where the latter was considered in \cite{Zhuang}.
Numerical results validate the advantages of the proposed technique
with respect to the ILM strategy of \cite{Zhuang}\cite{Schreck}.

The rest of the paper is organized as follows. Sec. \ref{sec:System-Model}
presents the system model. Sec. \ref{sec:SILM} formulates the problem
and introduces the proposed algorithm. Evaluation of the performance
of the proposed algorithm is presented in Sec. \ref{sec:Numerical-Results}
via numerical results. Finally, we conclude with some remarks in Sec.
\ref{sec: conclusions}.

\textit{Notation: }Bold uppercase letters denote matrices and bold
lowercase letters denote column vectors. The notations $\mathbb{E}$
and $\mathbb{C}$ are the expectation operator and the complex field,
respectively. $\mathcal{CN}\left(\mathbf{\boldsymbol{\mu}},\mathbf{\boldsymbol{\varSigma}}\right)$
represents the circularly symmetric distribution with mean vector
$\boldsymbol{\mu}$ and covariance matrix $\boldsymbol{\varSigma}$.
$\mathrm{tr}\left(\mathbf{A}\right)$ denotes the trace of the matrix
$\mathbf{A}$ and $\mathbf{A}^{H}$ is the conjugate transpose of
matrix $\mathbf{A}$. $v_{min}^{b}\left(\mathbf{A}\right)$ and $v_{max}^{b}\left(\mathbf{A}\right)$
are the truncated unitary matrices that consist of the $b$ eigenvectors
corresponding to the $b$ smallest and $b$ largest eigenvalues of
the non-negative definite matrix $\mathbf{A}$, respectively. The
Frobenius norm of a matrix $\mathbf{A}$ is denoted as $||\mathbf{A}||_{F}$.
$\mathbf{I}$ represents the identity matrix. $\mathbf{A^{\bot}}$
represents a unitary matrix that spans the subspace orthogonal to
the column space of the unitary matrix $\mathbf{A}$.

\section{System Model\label{sec:System-Model}}

We study the multi-cell MIMO system shown in Fig. \ref{fig:sys_model},
in which a subset of $L_{u}$ cells operates in the uplink while the
remaining $L_{d}$ cells operate in the downlink. We will use the
subscripts '$u$' and '$d$' throughout to denote the uplink and downlink
cells, respectively. Each BS has $N_{b}$ transmit/ receive antennas
and each mobile user has $N_{m}$ receive/ transmit antennas in the
downlink/ uplink cells. There are $K$ users per cell. We emphasize
that it is straightforward to generalize the analysis to arbitrary
numbers of users per cell and antennas. 

We focus on a standard channel model with flat-fading MIMO channels
that remain constant throughout the transmission block. Starting with
the downlink cells, the signal $\mathbf{y}_{\alpha_{d}k}\in\mathbb{C}^{N_{m}\times1}$
received by the $k$th user in cell $\alpha_{d}\in\left\{ 1,...,L_{d}\right\} $
is then given by

\begin{eqnarray}
\mathbf{y}_{\alpha_{d}k} & = & \underbrace{\mathbf{H}_{\alpha_{d}k}^{\alpha_{d}}\mathbf{x}_{\alpha_{d}k}}_{\mbox{desired signal}}+\underbrace{\sum_{j=1,j\neq k}^{K}\mathbf{H}_{\alpha_{d}k}^{\alpha_{d}}\mathbf{x}_{\alpha_{d}j}}_{\mbox{intra-cell interference}}\nonumber \\
 &  & +\underbrace{\sum_{\alpha_{u}=1}^{L_{u}}\sum_{j=1}^{K}\mathbf{H}_{\alpha_{d}k}^{\alpha_{u}j}\mathbf{x}_{\alpha_{u}j}}_{\mathrm{\mbox{uplink inter-cell interference}}}\nonumber \\
 &  & +\underbrace{\sum_{\beta_{d}=1,\beta_{d}\neq\alpha_{d}}^{L_{d}}\sum_{j=1}^{K}\mathbf{H}_{\alpha_{d}k}^{\beta_{d}}\mathbf{x}_{\beta_{d}j}}_{\mathrm{\mbox{downlink inter-cell interference}}}+\mathbf{n}_{\alpha_{d}k},\label{eq:received_signal-1-2}
\end{eqnarray}
where $\mathbf{H}_{\alpha_{d}k}^{\beta_{d}}$ is the $N_{m}\times N_{b}$
channel matrix from BS $\beta_{d}$ to user $k$ in cell $\alpha_{d}$,
where $\alpha_{d},\beta_{d}\in\left\{ 1,\cdots,L_{d}\right\} $; $\mathbf{x}_{\beta_{d}j}\in\mathbb{C}^{N_{b}\times1}$
is the transmitted signal vector from BS $\beta_{d}$ intended to
user $j$; $\mathbf{H}_{\alpha_{d}k}^{\alpha_{u}j}$ is the $N_{m}\times N_{m}$
channel matrix from user $j$ in cell $\alpha_{u}$ to user $k$ in
cell $\alpha_{d}$ and $\mathbf{n}_{\alpha_{d}k}$ denotes the thermal
noise at the considered user, which is assumed to be distributed as
$\mathcal{CN\mathrm{\mathbf{\mathrm{(\mathbf{0,I}).}}}}$ The first
term on the right-hand side of the equality is the desired signal,
the second term is the intra-cell interference, the third term is
the uplink inter-cell interference from all the users in the $L_{u}$
uplink cells and the forth term is the downlink inter-cell interference
from all the BSs in the $L_{d}$ downlink cells other than BS $\alpha_{d}$.

As for the uplink cells, the signal $\mathbf{y}_{\alpha_{u}}\in\mathbb{C}^{N_{b}\times1}$
received by BS $\alpha_{u}\in\left\{ 1,\cdots,L_{u}\right\} $ is
given by

\begin{eqnarray}
\mathbf{y}_{\alpha_{u}} & = & \underbrace{\sum_{k=1}^{K}\mathbf{H}_{\alpha_{u}}^{\alpha_{u}k}\mathbf{x}_{\alpha_{u}k}}_{\mbox{desired signal}}+\underbrace{\sum_{\beta_{u}=1,\beta_{u}\neq\alpha_{u}}^{L_{u}}\sum_{k=1}^{K}\mathbf{H}_{\alpha_{u}}^{\beta_{u}k}\mathbf{x}_{\beta_{u}k}}_{\mathrm{\mbox{uplink inter-cell interference}}}\nonumber \\
 &  & +\underbrace{\sum_{\alpha_{d}=1}^{L_{d}}\sum_{j=1}^{K}\mathbf{H}_{\alpha_{u}}^{\alpha_{d}}\mathbf{x}_{\alpha_{d}j}}_{\mathrm{\mbox{downlink inter-cell interference}}}+\mathbf{n}_{\alpha_{u}},\label{eq:received_signal-1-1-1}
\end{eqnarray}
where $\mathbf{H}_{\alpha_{u}}^{\beta_{u}k}$ is the $N_{m}\times N_{b}$
channel matrix from user $k$ in cell $\beta_{u}$ to BS $\alpha_{u}$,
where $\alpha_{u},\beta_{u}\in\left\{ 1,\cdots,L_{u}\right\} $; $\mathbf{x}_{\beta_{u}k}\in\mathbb{C}^{N_{m}\times1}$
is the transmitted signal vector from user $k$ in cell $\beta_{u}$;
$\mathbf{H}_{\alpha_{u}}^{\alpha_{d}}$ is the $N_{b}\times N_{b}$
channel matrix from BS $\alpha_{d}$ to BS $\alpha_{u}$ and $\mathbf{n}_{\alpha_{u}}$
denotes the thermal noise at the considered BS, which is assumed to
be distributed as $\mathcal{CN\mathrm{\mathbf{\mathrm{(\mathbf{0,I}).}}}}$
The first and second terms on the right-hand side of the equality
are interpreted as in (\ref{eq:received_signal-1-2}). The third term
is the downlink inter-cell interference from all the BSs in the $L_{d}$
downlink cells.

We assume that, in the downlink, we have the power constraint
\begin{equation}
\sum_{k=1}^{K}\mathbb{E}\left[\left\Vert \mathbf{x}_{\alpha_{\mathit{d}}k}\right\Vert ^{2}\right]=P\label{eq:power_constraint}
\end{equation}
for all $\alpha_{d}\in\left\{ 1,\cdots,L_{d}\right\} $. In the uplink,
equal power is used by all users yielding
\begin{equation}
\mathbb{E}\left[\left\Vert \mathbf{x}_{\alpha_{\mathit{u}}k}\right\Vert ^{2}\right]=P/K,\label{eq:power_constraint-2}
\end{equation}
for all $\alpha_{u}\in\left\{ 1,\cdots,L_{u}\right\} $ and $\textrm{ }k\in\left\{ 1,\cdots,K\right\} $.

In the downlink cells, for the transmission from BS $\alpha_{d}$
to user $k$, BS $\alpha_{d}$ chooses a unitary precoding matrix
$\mathbf{V}_{\alpha_{d}k}\in\mathbb{C}^{N_{b}\times s}$, where $s$
is the number of data streams per user. We assume throughout that
the number of data streams does not exceed the number of receive antennas
at each user, i.e., $s\leq N_{m}$, and that the total number of data
streams at each BS does not exceed the number of transmit antennas,
i.e., $Ks\leq N_{b}$. Moreover, for each user $k$ in the downlink
cell $\alpha_{d}$, a unitary matrix $\mathbf{G}_{\alpha_{d}k}\in\mathbb{C}^{N_{m}\times(N_{m}-s)}$
is selected that defines its \emph{interference subspace} as in \cite{Kumar}.
This is in the sense that the user pre-processes the received signal
as $\left(\mathbf{G}_{\alpha_{d}k}^{\bot}\right)^{H}\mathbf{y}_{\alpha_{d}k}$,
hence filtering out the received signal component that lie within
the interference subspace. We refer to the column space spanned by
$\mathbf{G}_{\alpha_{d}k}^{\bot}$ as the \emph{receive subspace}
for user $k$ in cell $\alpha_{d}$.

In the uplink cells, for the transmission from user $k$ to BS $\alpha_{u}$,
user $k$ chooses a precoding matrix $\mathbf{G}_{\alpha_{u}k}^{\bot}\in\mathbb{C}^{N_{m}\times s}$,
where $s$ is the number of data streams sent by each user. As in
the downlink case, we assume that the number of data streams satisfies
$s\leq N_{m}$ and $Ks\leq N_{b}.$ For each BS $\alpha_{u}$, a unitary
matrix $\mathbf{V}_{\alpha_{u}}^{\mathbf{}}\in\mathbb{C}^{N_{m}\times Ks}$
is used as the \emph{receiving subspace. }We observe that the notation
for the uplink cells is selected so as to be consistent with that
of the downlink cells.

In the next section, we will discuss how to design downlink matrices
$\mathbf{V}_{\alpha_{d}k}\in\mathbb{C}^{N_{b}\times s}$ and $\mathbf{G}_{\alpha_{d}k}\in\mathbb{C}^{N_{m}\times(N_{m}-s)}$
for all $\alpha_{d}\in\left\{ 1,\cdots,L_{d}\right\} $ and $\textrm{ }k\in\left\{ 1,\cdots,K\right\} $
and the uplink matrices $\mathbf{V}_{\alpha_{u}}\in\mathbb{C}^{N_{b}\times s}$
and $\mathbf{G}_{\alpha_{u}k}\in\mathbb{C}^{N_{m}\times(N_{m}-s)}$
for all $\alpha_{u}\in\left\{ 1,\cdots,L_{u}\right\} $ and $\textrm{ }k\in\left\{ 1,\cdots,K\right\} $.

\section{Signal and Interference Leakage Minimization\label{sec:SILM}}

Reference \cite{Kumar} proposed the SILM scheme for the design of
the precoding and receiving matrices for the special case $K=1$,
i.e., for a MIMO interference channel. Note that, in this case, we
can set $L_{u}=0$ or $L_{d}=0$ with no loss of generality. The SILM
scheme aims at striking a balance between two objectives: 1) minimizing
the interference power received by the users and BSs in the corresponding
receiving subspaces; 2) minimizing the signal power that is wasted
in the corresponding interference subspaces. This is done by adopting
as the optimization criterion the weighted sum of the power of the
interference leaked in the receive subspace and the power of the signal
wasted in the interference subspace. Specifically, an alternating
optimization method is proposed in which the precoding and receiving
matrices are optimized iteratively until convergence to a local minimum
of the performance criterion.

The proposed SILM scheme for the multicell uplink-downlink MIMO scenario
at hand combines the idea of SILM with the two-step precoding approach
for downlink channels studied in \cite{Zhuang}\cite{Schreck}\cite{Tse}.
Specifically, the overall precoding matrix $\mathbf{V}_{\alpha_{d}}=\left[\mathbf{V}_{\alpha_{d}1},\mathbf{V}_{\alpha_{d}2},\ldots,\mathbf{V}_{\alpha_{d}K}\right]$
used at the BS $\alpha_{d}$ is written as the product 
\begin{equation}
\mathbf{V}_{\alpha_{d}}=\mathbf{V_{\alpha_{\mathit{d}}}^{'}}\mathbf{V_{\alpha_{\mathit{d}}}^{''}},
\end{equation}
where $\mathbf{V_{\alpha_{\mathit{d}}}^{'}}=[\mathbf{V_{\alpha_{\mathit{d}}\mathrm{1}}^{'}},\cdots,\mathbf{V_{\alpha_{\mathit{d}}\mathrm{\mathit{K}}}^{'}}]\in\mathbb{C}^{N_{b}\times Ks},$
with $\mathbf{V}_{\alpha_{d}k}^{'}\in\mathbb{C}^{N_{b}\times s}$,
is a unitary matrix that is designed to handle \emph{uplink and downlink
inter-cell interference}, while $\mathbf{V_{\alpha_{\mathit{d}}}^{''}}=[\mathbf{V_{\alpha_{\mathit{d}}\mathrm{1}}^{''}},\cdots,\mathbf{V_{\alpha_{\mathit{d}}\mathrm{\mathit{K}}}^{''}}]\in\mathbb{C}^{Ks\times Ks}$,
with $\mathbf{V_{\alpha_{\mathit{d}}\mathrm{\mathit{k}}}^{''}}\in\mathbb{C}^{Ks\times s},$
is used to mitigate \emph{intra-cell interference}. In the following,
we discuss the design of the unitary downlink matrices $\mathbf{V}_{\alpha_{d}}^{\mathbf{'}}$
and $\mathbf{G}_{\alpha_{d}k},$ for all $\alpha_{d}\in\left\{ 1,\cdots,L_{d}\right\} $
and $k\in\left\{ 1,\cdots,K\right\} $ along with the unitary uplink
matrices $\mathbf{V}_{\alpha_{u}}^{\mathbf{'}}$ and $\mathbf{G}_{\alpha_{u}k},$
for all $\alpha_{u}\in\left\{ 1,\cdots,L_{u}\right\} $ and $\textrm{ }k\in\left\{ 1,\cdots,K\right\} $,
with the aim of handling inter-cell interference. We then detail the
calculation of the intra-cell precoding matrices $\mathbf{V_{\mathbf{\alpha_{\mathit{d}}}}^{''}}$.

\subsection{Uplink-Downlink Inter-Cell Precoding/Equalization}

In order to design the precoding and decoding matrices mentioned above,
we propose to minimize the sum 
\begin{equation}
\sum_{\alpha_{d}=1}^{L_{d}}\sum_{k=1}^{K}I_{\alpha_{d}k}+\sum_{\alpha_{u}=1}^{L_{u}}\sum_{k=1}^{K}I_{\alpha_{u}k},\label{eq:I-1}
\end{equation}
where $I_{\alpha_{d}k}$ is in turn defined as the weighted sum

\begin{eqnarray}
I_{\alpha_{d}k} & = & \sum_{\beta_{d}=1,\beta_{d}\neq\alpha_{d}}^{L_{d}}\left\Vert \left(\mathbf{G}_{\alpha_{d}k}^{\bot}\right)^{H}\mathbf{H}_{\alpha_{d}k}^{\beta_{d}}\mathbf{V}_{\beta_{d}}^{\mathbf{'}}\right\Vert _{F}^{2}\nonumber \\
 &  & +\sum_{\alpha_{u}=1}^{L_{u}}\sum_{j=1}^{K}\left\Vert \left(\mathbf{G}_{\alpha_{d}k}^{\bot}\right)^{H}\mathbf{H}_{\alpha_{d}k}^{\alpha_{u}j}\mathbf{G}_{\alpha_{u}j}^{\bot}\right\Vert _{F}^{2}\nonumber \\
 &  & +w\left\Vert \mathbf{G}_{\alpha_{d}k}^{H}\mathbf{H}_{\alpha_{d}k}^{\alpha_{d}}\mathbf{V}_{\alpha_{d}}^{\mathbf{'}}\right\Vert _{F}^{2},\label{eq:obj-1}
\end{eqnarray}
with $w\geq0$ being a given weight, and $I_{\alpha_{u}k}$ is defined
as the weighted sum

\begin{eqnarray}
I_{\alpha_{u}k} & = & \sum_{\beta_{u}=1,\beta_{u}\neq\alpha_{u}}^{L_{u}}\left\Vert \mathbf{V}_{\alpha_{u}}^{\mathbf{}H}\mathbf{H}_{\alpha_{u}}^{\beta_{u}k}\mathbf{G}_{\beta_{u}k}^{\bot}\right\Vert _{F}^{2}\nonumber \\
 &  & +\sum_{\alpha_{d}=1}^{L_{d}}\sum_{j=1}^{K}\left\Vert \mathbf{V}_{\alpha_{u}}^{\mathbf{}H}\mathbf{H}_{\alpha_{u}}^{\alpha_{d}}\mathbf{V}_{\alpha_{d}}^{\mathbf{'}}\right\Vert _{F}^{2}\nonumber \\
 &  & +w\left\Vert \mathbf{V}_{\alpha_{u}}^{\mathbf{}H}\mathbf{H}_{\alpha_{u}}^{\alpha_{u}k}\mathbf{G}_{\alpha_{u}k}^{\bot}\right\Vert _{F}^{2}.\label{eq:obj-1-1}
\end{eqnarray}
The expression (\ref{eq:obj-1}) is the sum in order of appearance,
of the inter-cell downlink and uplink interference powers (assuming
$\mathbf{V_{\alpha_{\mathit{d}}}^{''}}=\left(P/K\right)\mathbf{I}$)
and of the signal power wasted in the interference subspace for user
$k$ in cell $\alpha_{u}$, where the latter term is weighted by $w$,
and (\ref{eq:obj-1-1}) has a similar interpretation.

The optimization of (\ref{eq:obj-1}) is performed by alternating
between the minimization over the receive-side matrices $\mathbf{G}_{\alpha_{d}k}$
and $\mathbf{V}_{\alpha_{u}}^{\mathbf{}}$ for fixed transmit-side
matrices $\mathbf{V}_{\alpha_{d}}^{\mathbf{'}}$ and $\mathbf{G}_{\alpha_{u}j}$
and the minimization over the transmit-side matrices $\mathbf{G}_{\alpha_{u}j}$
and $\mathbf{V}_{\alpha_{d}}^{\mathbf{'}}$ for fixed receive-side
matrices $\mathbf{G}_{\alpha_{d}k}$ and $\mathbf{V}_{\alpha_{u}}^{\mathbf{}}$
following the procedure described in Table Algorithm \ref{alg:SILM-1}.
Specifically, in Step 2, the decoding matrices $\mathbf{G}_{\alpha_{d}\mathit{k}}^{\bot}$
and $\mathbf{V}_{\alpha_{u}}^{\mathbf{}}$ are obtained as

\begin{equation}
\mathbf{G}_{\alpha_{d}k}=v_{max}^{\left(N_{m}-s\right)}\left(\mathbf{\overrightarrow{\mathbf{Q}}}_{\alpha_{d}k}\right),\label{eq:Gak-1}
\end{equation}

\begin{equation}
\mathbf{V}_{\alpha_{u}}^{\mathbf{}}=v_{min}^{Ks}\left(\mathbf{\overleftarrow{\mathbf{Q}}}_{\alpha_{u}}\right),\label{eq:V_a-1-1-1}
\end{equation}
where
\begin{eqnarray}
\mathbf{\overrightarrow{\mathbf{Q}}}_{\alpha_{d}k} & = & \sum_{\beta_{d}=1,\beta_{d}\neq\alpha_{d}}^{L_{d}}\mathbf{H_{\alpha_{\mathit{d}}\mathit{k}}^{\mathit{\beta}_{\mathit{d}}}}\mathbf{V}_{\beta_{d}}^{\mathbf{'}}\mathbf{V}_{\beta_{d}}^{\mathbf{'}H}\mathbf{H}_{\alpha_{d}k}^{\beta_{d}H}\label{eq:Qak-1}\\
 &  & +\sum_{\alpha_{u}=1}^{L_{u}}\sum_{j=1}^{K}\mathbf{H}_{\alpha_{d}k}^{\alpha_{u}j}\mathbf{G}_{\alpha_{u}j}^{\bot}\left(\mathbf{G}_{\alpha_{u}j}^{\bot}\right)^{H}\mathbf{H}_{\alpha_{d}k}^{\alpha_{u}jH}\nonumber \\
 &  & -w\mathbf{H_{\alpha_{\mathit{d}}\mathbf{\mathrm{\mathit{k}}}}^{\mathit{\alpha_{d}}}V_{\alpha_{\mathit{d}}}^{'}V_{\alpha_{\mathit{d}}}^{'\mathit{H}}H_{\alpha_{\mathit{d}}\mathbf{\mathrm{\mathit{k}}}}^{\mathit{\alpha_{d}H}}}\nonumber 
\end{eqnarray}
and

\begin{eqnarray}
\mathbf{\overleftarrow{\mathbf{Q}}}_{\alpha_{u}} & = & \sum_{\beta_{u}=1,\beta_{u}\neq\alpha_{u}}^{L_{u}}\sum_{j=1}^{K}\mathbf{H}_{\alpha_{u}}^{\beta_{u}jH}\mathbf{G}_{\beta_{u}j}\mathbf{G}_{\beta_{u}j}^{H}\mathbf{H}_{\alpha_{u}}^{\beta_{u}j}\label{eq:Qa-1-1-1}\\
 &  & +\sum_{\beta_{u}=1}^{L_{u}}\mathbf{H}_{\alpha_{u}}^{\beta_{u}H}\mathbf{V}_{\beta_{u}}^{\mathbf{}}\mathbf{V}_{\beta_{u}}^{\mathbf{}H}\mathbf{H}_{\alpha_{u}}^{\beta_{u}}\nonumber \\
 &  & -w\sum_{j=1}^{K}\mathbf{H}_{\alpha_{u}}^{\alpha_{u}jH}\mathbf{G}_{\alpha_{u}j}\mathbf{G}_{\alpha_{u}j}^{H}\mathbf{H}_{\alpha_{u}}^{\alpha_{u}j}.\nonumber 
\end{eqnarray}
The first term on the right-hand side of (\ref{eq:Qak-1}) is the
covariance matrix of the downlink inter-cell interference at user
$k$ in cell $\alpha_{d}$; the second term is the covariance matrix
of the uplink inter-cell interference at user $k$ in cell $\alpha_{d}$;
and the third term is the weighted covariance matrix of the desired
signal for all the users in cell $\alpha_{d}$ as observed by user
$k$ in cell $\alpha_{d}$. The covariance matrix (\ref{eq:Qa-1-1-1})
has a similar interpretation.

In Step 3, the precoding matrices $\mathbf{V}_{\alpha_{d}}^{\mathbf{'}}$
and $\mathbf{G}_{\alpha_{u}k}^{\bot}$ are similarly obtained as 

\begin{equation}
\mathbf{V}_{\alpha_{d}}^{\mathbf{'}}=v_{min}^{Ks}\left(\mathbf{\overrightarrow{\mathbf{Q}}}_{\alpha_{d}}\right),\label{eq:V_a-1}
\end{equation}

\begin{equation}
\mathbf{G}_{\alpha_{u}k}=v_{max}^{\left(N_{m}-s\right)}\left(\mathbf{\overleftarrow{\mathbf{Q}}}_{\alpha_{u}k}\right),\label{eq:Gak-1-1-1}
\end{equation}
where

\begin{eqnarray}
\mathbf{\overrightarrow{\mathbf{Q}}}_{\alpha_{d}} & = & \sum_{\beta_{d}=1,\beta_{d}\neq\alpha_{d}}^{L_{d}}\sum_{j=1}^{K}\mathbf{H}_{\beta_{d}j}^{\alpha_{d}H}\mathbf{G}_{\beta_{d}j}\mathbf{G}_{\beta_{d}j}^{H}\mathbf{H}_{\beta_{d}j}^{\alpha_{d}}\nonumber \\
 &  & +\sum_{\alpha_{u}=1}^{L_{u}}\sum_{j=1}^{K}\mathbf{H}_{\alpha_{u}j}^{\alpha_{d}H}\mathbf{G}_{\alpha_{u}j}\mathbf{G}_{\alpha_{u}j}^{H}\mathbf{H}_{\alpha_{u}j}^{\alpha_{d}}\nonumber \\
 &  & -w\sum_{j=1}^{K}\mathbf{H}_{\alpha_{d}j}^{\alpha_{d}H}\mathbf{G}_{\alpha_{d}j}\mathbf{G}_{\alpha_{d}j}^{H}\mathbf{H}_{\alpha_{d}j}^{\alpha_{d}}\label{eq:Qa-1}
\end{eqnarray}
and 

\begin{eqnarray}
\mathbf{\overleftarrow{\mathbf{Q}}}_{\alpha_{u}k} & =\sum_{\alpha_{d}=1}^{L_{d}} & \sum_{j=1}^{K}\mathbf{H}_{\alpha_{d}j}^{\alpha_{u}kH}\mathbf{G}_{\alpha_{d}j}\mathbf{G}_{\alpha_{d}j}^{H}\mathbf{H}_{\alpha_{d}j}^{\alpha_{u}k}\label{eq:Qak-1-1-1}\\
 &  & +\sum_{\beta_{u}=1,\beta_{u}\neq\alpha_{u}}^{L_{u}}\sum_{j=1}^{K}\mathbf{H}_{\beta_{u}}^{\alpha_{u}kH}\mathbf{V}_{\beta_{u}}\mathbf{V}_{\beta_{u}}^{\mathbf{}H}\mathbf{H}_{\beta_{u}}^{\alpha_{u}k}\nonumber \\
 &  & -w\sum_{j=1}^{K}\mathbf{H}_{\alpha_{u}}^{\alpha_{u}kH}\mathbf{V}_{\alpha_{u}}^{\mathbf{}}\mathbf{V}_{\alpha_{u}}^{\mathbf{}H}\mathbf{H}_{\alpha_{u}}^{\alpha_{u}k}.\nonumber 
\end{eqnarray}
The first term on the right-hand side of (\ref{eq:Qa-1}) is the covariance
matrix of the downlink inter-cell interference caused by BS $\alpha_{d}$
to all downlink users; the second term is the covariance matrix of
the downlink inter-cell interference caused by BS $\alpha_{d}$ to
all the uplink BSs; and the third term is weighted covariance matrix
of the desired signal to all the users in cell $\alpha_{d}$ that
is leaked in the interference subspaces. The covariance matrix (\ref{eq:Qak-1-1-1})
has a similar interpretation.

\begin{algorithm}[t]
\protect\caption{\label{alg:SILM-1}Signal and Interference Leakage Minimization (SILM)
for the multicell MIMO downlink}

\textbf{Step 1:} Start with arbitrary unitary precoding matrices $\mathbf{V}_{\alpha_{d}}^{\mathbf{'}}$
for all $\alpha_{d}\in\left\{ 1,\cdots,L_{d}\right\} $ and $\mathbf{G}_{\alpha_{\mathit{u}}k}$
for all $\alpha_{u}\in\left\{ 1,\cdots,L_{u}\right\} $. 

\textbf{Step 2:} Compute $\mathbf{G}_{\alpha_{d}\mathit{k}}$ as in
(\ref{eq:Gak-1}) for all $\alpha_{d}\in\left\{ 1,\cdots,L_{d}\right\} $
and $k\in\left\{ 1,\cdots,K\right\} $ and $\mathbf{V_{\alpha_{\mathit{u}}}}$
as in (\ref{eq:V_a-1-1-1}) for all $\alpha_{u}\in\left\{ 1,\cdots,L_{u}\right\} $.

\textbf{Step 3:} Compute $\mathbf{V_{\alpha_{\mathit{d}}}^{'}}$ as
in (\ref{eq:V_a-1}) for all $\alpha_{d}\in\left\{ 1,\cdots,L_{d}\right\} $
and $\mathbf{G}_{\alpha_{u}k}$ as in (\ref{eq:Gak-1-1-1}) for all
$\alpha_{u}\in\left\{ 1,\cdots,L_{u}\right\} $ and $k\in\left\{ 1,\cdots,K\right\} $.

\textbf{Step 4: }If a convergence criterion is satisfied, go to Step
5; otherwise go back to Step 2.

\textbf{Step 5: }Compute $\mathbf{V_{\mathbf{\alpha_{\mathit{d}}}}^{''}}$
using (\ref{eq:V_prime}).
\end{algorithm}

\textbf{\textit{Remark 1: }}The alternating optimizations algorithm
in Table Algorithm \ref{alg:SILM-1} can be implemented in a distributed
fashion where Step 2 is carried out in parallel by all the downlink
users and uplink BSs, while Step 3 is performed in parallel by all
the downlink BSs and uplink users. In both steps, only local channel
state information is needed if the users and BSs, e.g., BS $\alpha_{d}$
only needs to know the channels $\mathbf{H}_{\beta_{d}k}^{\alpha_{d}}$
for all $\beta_{d}\in\left\{ 1,\cdots,L_{d}\right\} $ and $k\in\left\{ 1,\cdots,K\right\} $.
Moreover, BSs and users need to exchange unitary matrices during the
operation of the algorithm in order to calculate the covariance matrices
(\ref{eq:Qak-1}), (\ref{eq:Qa-1-1-1}), (\ref{eq:Qa-1}) and (\ref{eq:Qak-1-1-1}).\hfill{}$\square$

Given the matrices $\mathbf{G}_{\alpha_{d}k}$ and $\mathbf{V}_{\alpha_{d}}^{\mathbf{'}}$,
the effective channel observed by the $K$ users in the downlink cell
$\alpha_{d}$ from BS $\alpha_{d}$ is given as

\begin{equation}
\tilde{\mathbf{H}}_{\alpha_{d}}=\left[\left(\mathbf{G}_{\alpha_{d}k}^{\bot}\right)^{H}\mathbf{H}_{\alpha_{d}1}^{\alpha_{d}}\mathbf{V_{\alpha_{\mathit{d}}}^{'}};\cdots;\left(\mathbf{G}_{\alpha_{d}k}^{\bot}\right)^{H}\mathbf{H}_{\alpha_{d}K}^{\alpha_{d}}\mathbf{V_{\alpha_{\mathit{d}}}^{'}}\right].\label{eq:H_prime}
\end{equation}
Each BS $\alpha_{d}$ then precodes over this channel so as to control
intra-cell interference. Here, we adopt a linear MMSE intra-cell precoder,
which is given as 
\begin{equation}
\mathbf{V_{\mathbf{\alpha_{\mathit{d}}}}^{''}}=\left(\mu_{\alpha_{d}}\mathbf{I}+\tilde{\mathbf{H}}_{\alpha_{d}}\tilde{\mathbf{H}}_{\alpha_{d}}^{H}\right)^{-1}\tilde{\mathbf{H}}_{\alpha_{d}}^{H},\label{eq:V_prime}
\end{equation}
where $\mu_{\alpha_{d}}$ is a scalar that must be selected such that
$\mathrm{tr}[\mathbf{V_{\mathbf{\alpha_{\mathit{d}}}}}\mathbf{V}_{\alpha_{d}}^{\mathbf{}H}]=P$.
Note that with $\mu_{\alpha_{d}}=0$, the MMSE solution in (\ref{eq:V_prime})
reduces to the ZF design considered in \cite{Zhuang}.

\subsection{Sum-rate\label{sec:rate-1}}

\begin{figure*}[t]
\begin{eqnarray}
R^{DL} & = & \sum_{\alpha_{d}=1}^{L_{d}}\sum_{k=1}^{K}\mathrm{log}\left|\mathbf{I}+\left(\mathbf{I}+\left(\mathbf{G}_{\alpha_{d}k}^{\bot}\right)^{H}\right.\right.\nonumber \\
 &  & \left(\overset{}{\underset{i=1,i\neq k}{\sum^{K}}}\mathbf{H}_{\alpha_{d}k}^{\alpha_{d}H}\mathbf{V}_{\alpha_{d}i}\mathbf{V}_{\alpha_{d}i}^{\mathit{H}}\mathbf{H}_{\alpha_{d}\mathbf{\mathrm{\mathit{k}}}}^{\mathit{\alpha}}+\overset{}{\underset{\beta_{d}=1,\beta_{d}\neq\alpha_{d}}{\sum^{L_{d}}}}\mathbf{H}_{\alpha_{d}k}^{\beta_{d}}\mathbf{V}_{\beta_{d}}\mathbf{V}_{\beta_{d}}^{\mathit{H}}\mathbf{H}_{\alpha_{d}k}^{xH}\right.\nonumber \\
 &  & \left.\left.\left.+\frac{P}{Kd}\sum_{\alpha_{u}=1}^{L_{u}}\sum_{j=1}^{K}\mathbf{H}_{\alpha_{d}k}^{\alpha_{u}j}\mathbf{G}_{\alpha_{u}j}^{\bot}\left(\mathbf{G}_{\alpha_{u}j}^{\bot}\right)^{H}\mathbf{H}_{\alpha_{d}k}^{\alpha_{u}jH}\right)\mathbf{G}_{\alpha_{d}k}^{\bot}\right)^{-1}\left(\mathbf{G}_{\alpha_{d}k}^{\bot}\right)^{H}\mathbf{H}_{\alpha_{d}k}^{\alpha_{d}}\mathbf{V}_{\alpha_{d}\mathbf{\mathrm{\mathit{k}}}}\mathbf{V}_{\alpha_{d}\mathbf{\mathrm{\mathit{k}}}}^{\mathit{H}}\mathbf{H}_{\alpha_{d}\mathbf{\mathrm{\mathit{k}}}}^{\mathit{\alpha_{d}H}}\mathbf{G}_{\alpha_{d}k}^{\bot}\right|.\label{eq:BC_rate-1-1}
\end{eqnarray}
\end{figure*}

\begin{figure*}[t]
\begin{eqnarray}
R^{UL} & = & \sum_{\alpha_{u}=1}^{L_{u}}\mathrm{log\left|\mathit{\mathbf{I}+\frac{P}{Kd}\left(\mathbf{I}+\mathbf{V}_{\alpha_{u}\mathbf{\mathrm{\mathit{}}}}^{\mathbf{}\mathit{H}}\right.}\right..}\nonumber \\
 &  & \left.\left(\frac{P}{Kd}\overset{L_{u}}{\underset{\beta_{u}=\mathrm{1},\beta_{u}\neq\alpha_{u}}{\sum}}\overset{K}{\underset{i=\mathrm{1}}{\sum}}\left(\mathbf{H}_{\beta_{u}}^{\mathit{\alpha_{u}k}}\right)^{H}\mathbf{G}_{\beta_{u}i}^{\bot}\left(\mathbf{G}_{\beta_{u}i}^{\bot}\right)^{H}\mathbf{H}_{\beta_{u}}^{\mathit{\alpha_{u}k}}+\sum_{\alpha_{d}=1}^{L_{d}}\sum_{j=1}^{K}\mathbf{H}_{\alpha_{d}j}^{\alpha_{u}kH}\mathbf{V}_{\alpha_{d}j}\mathbf{V}_{\alpha_{d}j}^{H}\mathbf{H}_{\alpha_{d}j}^{\alpha_{u}k}\right)\mathbf{V}_{\alpha_{u}}^{\mathbf{}}\right)^{-1}\nonumber \\
 &  & \left.\mathbf{V}_{\alpha_{u}}^{\mathbf{}\mathit{H}}\overset{K}{\underset{k=1}{\sum}}\left(\mathbf{H}_{\alpha_{u}}^{\mathit{\alpha_{u}k}}\right)^{H}\mathbf{G}_{\alpha_{u}k}^{\bot}\left(\mathbf{G}_{\alpha_{u}k}^{\bot}\right)^{H}\mathbf{H}_{\alpha_{u}}^{\mathit{\alpha_{u}k}}\mathbf{V}_{\alpha_{u}\mathbf{\mathrm{\mathit{}}}}^{\mathbf{}}\right|.\label{eq:MAC_rate-1}
\end{eqnarray}

\_\_\_\_\_\_\_\_\_\_\_\_\_\_\_\_\_\_\_\_\_\_\_\_\_\_\_\_\_\_\_\_\_\_\_\_\_\_\_\_\_\_\_\_\_\_\_\_\_\_\_\_\_\_\_\_\_\_\_\_\_\_\_\_\_\_\_\_\_\_\_\_\_\_\_\_\_\_\_\_\_\_\_\_\_\_\_\_\_\_\_\_\_\_\_\_\_\_\_\_
\end{figure*}

Given the designed precoding and decoding matrices, assuming that
all interference is treated as noise, the sum-rate of the downlink
cells can be computed as in (\ref{eq:BC_rate-1-1}) and the the sum-rate
of the uplink cells can be obtained in a similar fashion as in (\ref{eq:MAC_rate-1}).
Note that the latter assumes equal power allocation per stream and
joint decoding of all the uplink users in a cell.

\section{Numerical Results\label{sec:Numerical-Results}}

In this section, we present some numerical results for the schemes
under study. We assume that all the channel matrices corresponding
to a BS and a user in the same cell are independent and identically
distributed (i.i.d.) as $\mathcal{CN\mathrm{\mathbf{\mathrm{(0,1)}}}}$
and all channel matrices corresponding to a BS and a user in a different
cell are i.i.d. as $\mathcal{CN}\mathrm{\mathbf{\mathrm{(0,\rho^{2})}}},$
where $\rho$ can be interpreted as the inter-cell interference gain.
We define the signal-to-noise ratio (SNR) as being equal to $P$.

\begin{figure}[t]
\centering\includegraphics[bb=0bp 4bp 600bp 400bp,clip,scale=0.5]{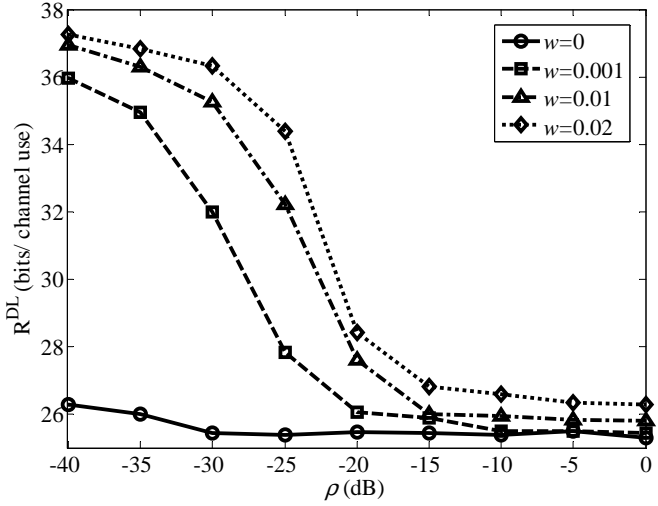}\protect\caption{\label{fig:BC} Sum-rate $R^{DL}$ for the downlink with MMSE precoding
versus the inter-cell interference gain $\rho$ ($L_{d}=4,L_{u}=0,K=4,N_{b}=N_{m}=5,s=1,SNR=10dB$).}
\end{figure}

We first consider the special case where all cells operate in the
downlink. Fig. \ref{fig:BC} plots the downlink sum-rate $R^{DL}$
for the downlink versus the inter-cell interference gain $\rho$ for
$L_{d}=4,L_{u}=0,K=4,N_{b}=N_{m}=5,s=1$ and $SNR=10dB$. The performance
of SILM, which corresponds to $w>0$, is shown for MMSE precoding.
As $\rho$ decreases, the performance of ILM, which corresponds to
$w=0$, is significantly degraded as compared to SILM since the contribution
of the inter-cell interference becomes relatively less relevant. It
is also noted that values of $w$ larger than $0.02$ do not further
improve the performance (not shown).

\begin{figure}[t]
\centering\includegraphics[bb=3bp 2bp 500bp 340bp,clip,scale=0.59]{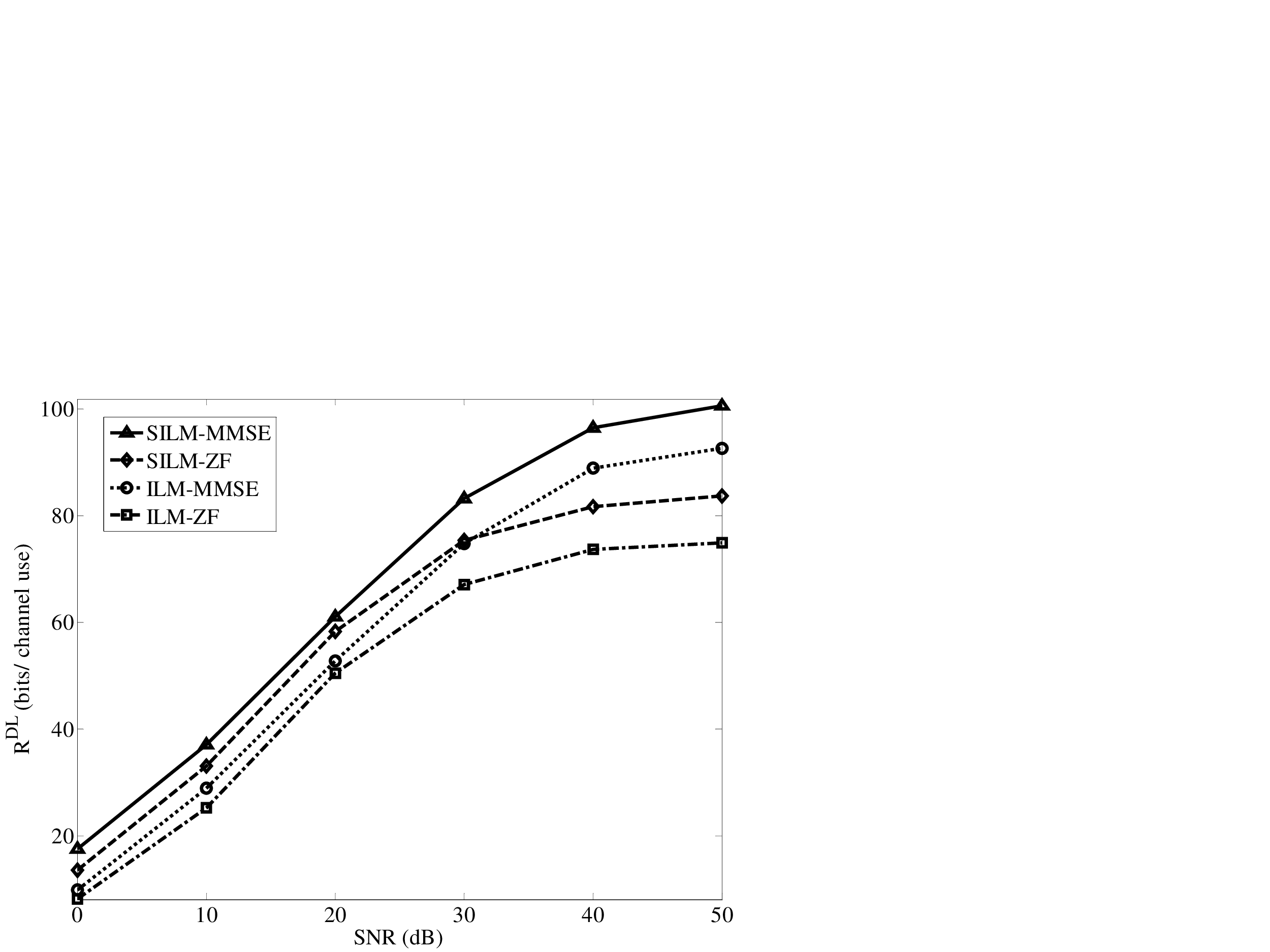}\protect\caption{\label{fig:gain} Sum-rate $R^{DL}$ for the downlink channel versus
SNR for ILM-ZF, SILM-ZF, ILM-MMSE and SILM-MMSE ($L_{d}=4,L_{u}=0,K=5,N_{b}=N_{m}=5,s=1,\rho=-20dB$).}
\end{figure}

Fig. \ref{fig:gain} plots the downlink sum-rate $R^{DL}$ for the
downlink versus the SNR for $L_{d}=4,L_{u}=0,K=5,N_{b}=N_{m}=5,s=1$
and $\rho=-20dB$. Following \cite{Kumar}, the weights $w$ corresponding
to the SNR values $\left[0\;10\;20\;30\;40\;50\right]$ are selected
as $\left[0.02\;0.02\;0.005\;0.003\;0.002\;0.001\right].$ Note that
the weight value decreases with the SNR, reflecting the enhanced role
of interference in the high-SNR regime. The performance of SILM with
MMSE and ZF intra-cell precoding is compared to the ILM scheme. It
is confirmed that SILM-based algorithms outperform ILM schemes. Moreover,
MMSE precoding significantly improves the sum-rate over ZF.

\begin{figure}[t]
\centering\includegraphics[bb=0bp 0bp 680bp 400bp,clip,scale=0.48]{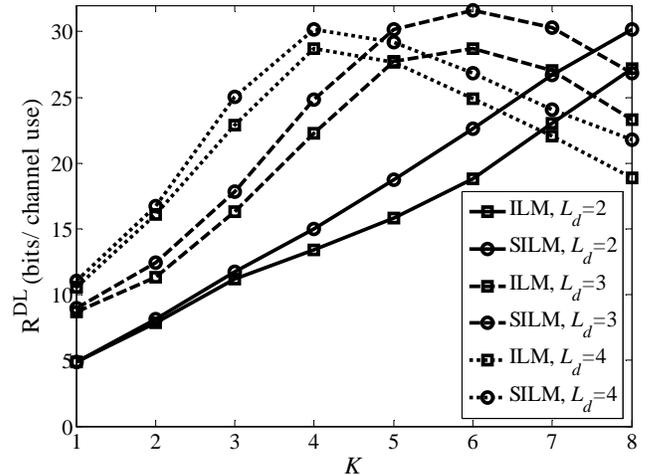}\protect\caption{\label{fig:mac} Sum-rate $R^{DL}$ for the downlink versus $K$ for
ILM and SILM ($L_{d}=2,3,4,L_{u}=0,N_{b}=N_{m}=5,s=1,\rho=-20dB$).}
\end{figure}

Fig. \ref{fig:mac} plots the sum-rate $R^{DL}$ versus the number
$K$ of users per cell for ILM and SILM for $L_{d}=2,3,4,L_{u}=0,N_{b}=N_{m}=5,s=1$
and $\rho=-20dB$. It is seen that the gain of SILM over ILM decreases
as the number of cells and/or of users per cell increases, and hence
as the performance of the system becomes increasingly interference
limited. Note also that there is an optimal number of users $K$ due
to the assumption of fixed power allocation.

\begin{figure}[tb]
\centering\includegraphics[bb=5bp 5bp 550bp 400bp,clip,scale=0.5]{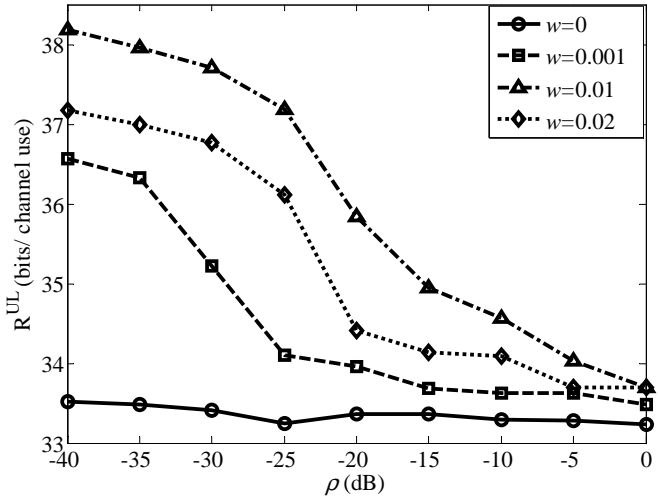}\protect\caption{\label{fig:fig4} Sum-rate $R^{UL}$ for the uplink versus $\rho$
($L_{d}=0,L_{u}=4,K=4,N_{b}=N_{m}=5,s=1,SNR=10dB$).}
\end{figure}

We now consider the case of all uplink cells. Fig. \ref{fig:fig4}
plots the sum-rate $R^{UL}$ for the uplink versus the inter-cell
interference gain $\rho$ for $L_{d}=0,L_{u}=4,K=4,N_{b}=N_{m}=5,s=1$
and $SNR=10dB$. As observed for the downlink in Fig. \ref{fig:BC},
SILM becomes more advantageous as $\rho$ decreases, and hence as
the performance becomes less limited by the inter-cell interference
as implicitly assumed by ILM. It is also noted that $w=0.02$ yields
lower performance than $w=0.01.$

\begin{figure}[t]
\centering\includegraphics[bb=0bp 0bp 580bp 400bp,clip,scale=0.48]{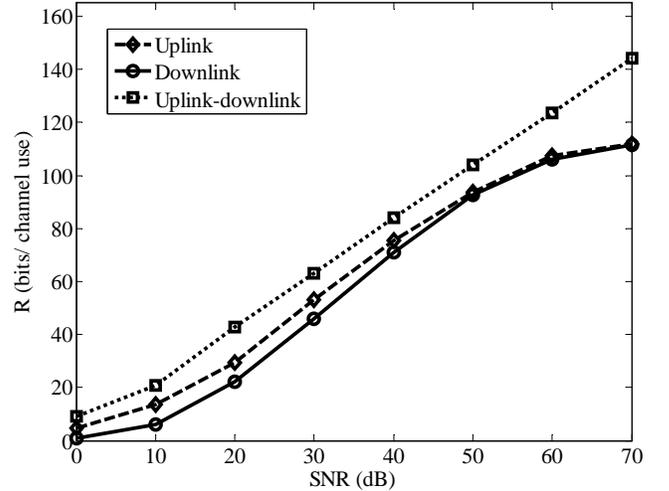}\protect\caption{\label{fig:fig4-1} Sum-rate $R$ for the uplink-downlink versus SNR
for SILM-MMSE ($L_{d}=2,L_{u}=2,K=2,N_{b}=N_{m}=4,s=1,\rho=-20dB$).}
\end{figure}

Finally, Fig. \ref{fig:fig4-1} plots the sum-rate versus the SNR
for a system with four cells. We compare the performance of the downlink
configuration with $L_{d}=4$ and $L_{u}=0$, of the uplink with $L_{d}=0$
and $L_{u}=4$ and of the uplink-downlink configuration with $L_{d}=2$
and $L_{u}=2$ $K=2,N_{b}=N_{m}=4,s=1$ and $\rho=-20dB$. The uplink
configuration outperforms the downlink solution due to the assumed
joint decoding of the intra-cell users that contrasts with the assumed
linear intra-cell downlink precoding. It is also observed that the
uplink-downlink configuration provides significant performance gains
especially at high SNR, confirming the results in \cite{Jeon}.

\section{Concluding Remarks\label{sec: conclusions}}

In this paper, a novel algorithm for the design of linear transmit-
and receive-side processing has been proposed for a MIMO multicell
system with different cells operating in either uplink and downlink.
The algorithm is based on the minimization of the weighted sum of
the interference power that is leaked outside the interference subspace
and of the signal power that falls into the interference subspace
as in \cite{Kumar}. The proposed technique generalizes the interference-leakage
based approach of \cite{Kumar}-\cite{Zhuang} and is shown via numerical
results to have significant sum-rate gains over existing techniques.

\end{document}